\documentclass[fleqn,usenatbib]{mnras}

\usepackage{newtxtext,newtxmath}

\usepackage[T1]{fontenc}

\DeclareRobustCommand{\VAN}[3]{#2}
\let\VANthebibliography\thebibliography
\def\thebibliography{\DeclareRobustCommand{\VAN}[3]{##3}\VANthebibliography}

\usepackage{graphicx}	
\usepackage{amsmath}	
\usepackage{array}

\newcolumntype{M}{>{\centering\arraybackslash}m{2cm}}
\newcolumntype{S}{>{\centering\arraybackslash}m{1cm}}
\newcolumntype{L}{>{\centering\arraybackslash}m{3cm}}
\newcolumntype{X}{>{\centering\arraybackslash}m{4cm}}

\title[Review of synergic meteor observations]{Review of synergic meteor observations: linking the results from cameras, ionosondes, infrasound and seismic detectors}

\author[A. Kereszturi et al.]{
Á. Kereszturi,$^{1,3}$\thanks{E-mail: kereszturi.akos@csfk.org}
V. Barta,$^{2}$
I. Bondár,$^{5}$
Cs. Czanik,$^{2}$
A. Igaz,$^{1,3}$
P. Mónus,$^{2}$
D. Rezes,$^{1,4}$
L. Szabados,$^{1}$
\newauthor and B. D. Pál$^{1}$
\\
$^{1}$Konkoly Observatory, Konkoly Thege Miklós Astronomical Institute, Research Centre for Astronomy and Earth Sciences, ELKH, Hungary\\
$^{2}$Earth Physics and Space Science Institute, ELKH, Hungary\\
$^{3}$Hungarian Astronomical Association, Hungary\\
$^{4}$ELTE Department of Petrology and Geochemistry, Hungary\\
$^{5}$ Institute for Geological and Geochemical Research, Research Centre for Astronomy and Earth Sciences, ELKH, Hungary
}

\date{Accepted XXX. Received YYY; in original form ZZZ}

\pubyear{2021}

\begin{document}
\label{firstpage}
\pagerange{\pageref{firstpage}--\pageref{lastpage}}
\maketitle

\begin{abstract}
Joint evaluation of different meteor observation types support the better understanding of both the meteor phenomenon and the terrestrial atmosphere. Two types of examples are presented in this work, linking ionospheric effects to specific meteors, where almost one third of the meteors emerged at high altitudes were simultaneously recorded with an optical camera. Very few such observations have been realized yet. With a daytime fireball, the recorded infrasound effect and the atmospheric blast produced shock wave related small earthquakes were identified by a network of ground stations. The overview of these observational types highlights specific topics where substantial improvements and discoveries are expected in the near future. 
\end{abstract}




\section{Introduction}

The aim of this work is to overview the potential of connecting different types of observations and aspects of meteor phenomena, specifically fusing optical camera based, ionosonde, infrasound and seismic detectors based results; while also providing an outlook for the future. Several fields of meteor observation were developed recently, including automatic sky monitoring camera systems \citep{toth2013,toth2014}, spectral recording of meteors during their flight \citep{ferus2020}, meteor phenomena recording based on radio \citep{sung2020} and radar \citep{chen2020} observing methods. The target of these studies are the impacting meteoritic bodies burning up in the atmosphere of the Earth, that occasionally hit the ground. In ideal cases, these could be found and recovered \citep{spurny2020} right after the infall, using the calculated trajectory from observations.\\

This work summarizes partly the recent works in the above listed topics in Hungary, where such activities are running simultaneously. Using these results and experiences, the possibility to build up an interconnected system is outlined, where data from different observational types could be linked and interpreted jointly. Their results will help to better understand not only the meteor phenomena but some characteristics of near Earth space, physical properties of meteoritic bodies, as well as our atmosphere. Although the currently existing system is not ready for an idealized high level synthesis, the results presented here provide the basis of planning and improving such joint evaluation of individual studies, with the identification of possible benefits and future scientific expectations. 

\section{Background information}

Meteor phenomenon occurs in the upper atmosphere of the Earth, roughly in the elevation between 60-90 km above the surface \citep{bariselli2020}. The body (a meteoroid) producing it ablates totally in most cases, however especially the larger ones could reach the surface in a dark flight after a decelerating phase. In exceptional cases, hypersonic impacts can happen producing classical impact craters. The atmospheric flight of a meteoroid has various consequences, influencing different atmospheric layers, producing thermal, optical, ionization, VLF wave radio emission, propagating blast waves, acoustic and seismic effects \citep{revelle2003} --– but the observations of all these effects require different recording methods. A common treatment of these different aspects is important not only to understand the runoff of the meteor phenomenon \citep{khrennikov2020,palotai2019}, but also to better estimate the current bombardment rate of near Earth space \citep{sekhar2014} and the related occurrence of small bodies. Their atmospheric ablation \citep{dias2020}, deceleration \citep{vojacek2019} and fragmentation \citep{spurny2020} give constraints for their internal structure \citep{wheeler2016}, which might be important for the geological history of parent bodies \citep{borovicka2019}, and even for suggesting mitigation actions \citep{nuth2017} against them. The understanding of atmospheric flight and the behavior of meteoroids support the successful recovery of fallen meteorites \citep{spurny2020}. \\
 
Upon its entry, the meteoroid has an influence on the upper atmosphere of the Earth. The so-called Sporadic E (Es) layers of the ionosphere are thin regions of enhanced electron density at the 80 - 150 km altitude region. Their origin is explained by the windshear theory at mid-latitudes as the long-living ions originating from meteors form thin layers owing to the shears of the tidal winds \citep{whitehead1989,haldoupis2011}. The link between meteor rates and Es layers has been broadly discussed in the literature. There is a similarity between the seasonal dependence of the detected meteor rates and the observed Sporadic E activity, which suggested a cause-and-effect relation between the two phenomena \citep{haldoupis2007}. A comparison between Es rates (measured by GPS radio occultation (RO)) and meteor rates (provided by VHF meteor radar) during the Geminid meteor showers has shown that Es activity increased after the shower with a few days delay in most cases between 2006 and 2010 \citep{jacobi2013}. \\

Furthermore, plasma traces of individual meteors have also been recognized in the ionograms only recently, as such observation of individual meteors by ionosondes is not an easy task. Generally speaking, the ionosondes detect ionograms every 15 mins in standard mode. The plasma trail of a meteor remains at the 80-120 km height for 1-1.5 mins or less based on atmospheric chemistry models and observations \citep{zinn2005}. Therefore, the chances are low to detect the plasma trail of individual meteors with an ionosonde. However it is possible to detect a trail with high cadence campaign measurements when ionograms are recorded in every minute. Firstly, \citet{goldsbrough1976} reported the observation of echoes \textbf{in} ionograms reflected from spontaneous, long, high-electron-line-density meteor trails as a class of Es layers during meteor showers. Other detected effects on Es activity \textbf{(as a consequence of meteor showers)} are the increases in sporadic E occurrence rate \citep{chandra2001}, Es ionization related to fireball events (exceptionally large meteor events, \citet{rajaram1991}) and delayed increases in Es activity \citep{sinno1980}. Furthermore, \citet{maruyama2003} and \citet{maruyama2008} made observations with rapid-run ionosondes during Leonids 2001 and Perseids 2002 meteor showers and reported long duration echoes\textbf{, which} they determined as the formation of meteor-induced sporadic E patches. In spite of the previous observations, there are still many unanswered questions in the relationship between meteor rates and Es activity: How do the different meteor showers affect the Sporadic E activity? Does the impact of the shower on the Es layers depend on the origin or the composition (concentration of the metallic ions)? How long can the traces of individual meteors or meteor-induced Es patches be observed in the ionograms? \\

Meteors traveling with supersonic speeds through the atmosphere generate infrasound as a line source. The direction of the infrasound wave propagation is perpendicular to the trajectory of the meteor, and thus only sensors favorably situated with respect to the trajectory of the meteor are able to record the infrasound signal \citep{pilger2015}. Bolides that catastrophically disrupt the atmosphere at the end of their trajectory (occasionally, but not properly called explosion) generate additional strong acoustic signals by the shock wave that produces the tell-tale N-shape signal of a point-source event on microbarometers \citep{edwards2010,ens2012,silber2019}. Since low-frequency infrasound waves can travel to large distances with little attenuation in the atmosphere, they often provide the only means to detect meteors and bolides entering the atmosphere over oceans and remote locations lacking local observing capacities. Seismometers with a clear line of sight often record acoustic signals generated by meteors (e.g. \citet{borovicka2013}; \citet{brown2002a}; \citet{groot2019}; \citet{hedlin2010}; \citet{spurny2010};  \citet{walker2010}). \\
 
The first infrasound recording of a Near Earth Object (NEO) was the Tunguska impact in 1908 \citep{whipple1930}. The characteristic N-shape signal of a sonic boom and the nearly negligible attenuation (that allows infrasound waves to travel thousands of kilometers in the atmosphere) made infrasound the primary technology to monitor atmospheric nuclear explosions during the Cold War. It was soon realized that some of the explosive-like events recorded by infrasound instruments were not associated with nuclear explosions at all, but the events were caused by bolides. This historical data set on large bolides provided the basis for the estimation of the influx of small-size NEOs \citep{revelle1997,revelle2008a,brown2002b,silber2009}. However, with the Limited Nuclear-Test-Ban Treaty that banned atmospheric and underwater nuclear tests in 1964, funding for infrasound research quickly dried up and the field almost became extinct. \\
 
The Comprehensive Nuclear-Test-Ban Treaty (CTBT) brought a renaissance to the discipline, as one of the four technologies for monitoring nuclear explosions is infrasound. The International Monitoring System (IMS) infrasound network consists of 60 arrays (distributed worldwide), and it is designed to detect a 1 kiloton nuclear explosion anywhere on the globe \citep{christie2010}. The IMS network regularly detects bolides, including one of the largest in \textbf{recorded} history, the Chelyabinsk fireball of 15 February 2013 \citep{brown2013,pichon2013,pilger2015,pilger2019}. The IMS network is complemented by regional infrasound networks such as the ARISE (Atmospheric dynamics Research InfraStructure in Europe) network \citep{pilger2018} and the Transportable Array in the US \citep{groot2019}. \\
 
Infrasound records provide an estimate for the kinetic energy of the meteor by the empirical period-yield relations derived from atmospheric nuclear explosions. For a network of infrasound stations, the mean signal period is used to estimate the total energy released by a bolide (e.g. \citet{silber2011, ens2012,brown2013}). Other types of observations like optical or radar might provide supporting information. Around the terminal disruption(s) only infrasound methods can show how many and what type of characteristic catastrophic fragmentations happened, as optical detectors become oversaturated by the terminal and most energetic disruptions. \\
 
All-sky camera systems co-located or working in a coordinated way with infrasound arrays have been proven to be particularly successful in detecting meteors and determining their physical parameters \citep{weryk2008,silber2014,silber2019}. We expect similar performance from the Piszkés-tető Observatory at the authors’ host institute, with its co-located all-sky camera, infrasound array and seismological station. The infrasound array has been operational since May 25, 2017. \\

Several topics could have been also linked to the above listed meteor related aspects (like radar, meteor spectra recording) but as no strong activity is present among the collaborators of this work, these are out of the scope of this paper. 

\section{Methods, Observations, Simulations}

Various meteor observing methods were used in this work, which all provide information on the meteor phenomenon. These are presented below to provide background and context for the interpretation of the observations, presenting ionospheric sounding, optical camera, infrasound detectors, and seismic detectors, in this order. 

\subsection{Optical observations}

The optical images provide the burn-up and flight track of meteors, which usually starts at about 70-100 km altitude, and might reach down to 30-40 km altitude, or even below in the case of large (m sized) bodies. This flight track is only recordable above a certain apparent brightness. This means, that the meteor might have already been burning in the atmosphere above the highest optically recorded elevation (where the brightness was not high enough yet) (see Sect. \ref{sec:ionosp}). This is also true for the lower part of the flight, when the decelerated body stops emitting light (this is the so-called dark flight part, only relevant for objects in the cm-dm scale). It is also worth noting that larger meteors can fragment catastrophically at the termination point at the end of their flight path, which is difficult to observe precisely as the optical detectors usually become oversaturated. \\

The optical data used in this work were recorded by a Watec 902H2 Ultimate camera, equipped with a Computar HG2610AFCS-HSP objective (with a focal length of 2.6 mm, 30 mm effective lens aperture, and 122 $\times$ 97 degree field of view toward the zenith.) This station operates with the Metrec \citep{molau1999} automatic meteor detection software. Its limiting magnitude is roughly +1 for meteors in this work.

\subsection{Seismological observation methods}

The first digital seismic station at Piszkés-tető became operational in 1991 \citep{bondar1992}. The Paks Microseismic Monitoring Network \citep{doi1995} was established to monitor the seismicity around the Paks nuclear power plant in 1995, and although it is privately operated, it formed the backbone of the Hungarian National Seismological Network (HNSN, \citet{doi1992}) until 2011. By then the HNSN consisted of 10 digital broadband stations. The Kövesligethy Radó Seismological Observatory (KRSZO) participated in the AlpArray \citep{doi2015} experiment \citep{hetenyi2018} with 11 temporary broadband stations \citep{graczer2018}. As of today, the HNSN consists of 15 permanent and 26 PAnnonian-Carpathian-Alpine Seismic Experiment (PACASE) temporary broadband stations. \\

There were 57 seismological stations working in Hungary in 2020. A part of them (41) has been operated by the former Seismological Observatory of former Geodetic and Geophysical Institute, Research Centre for Astronomy and Earth Sciences (CSFK GGI), the others by GeoRisk Ltd.. These latter 16 stations belong to the Paks Nuclear Power Plant (FDSN network code: HM). Stations operated by GGI can be divided into two groups: 15 stations form the permanent network of GGI (network code: HU), the other 26 stations are temporary stations financed by two projects: AlpArray and PACASE (network codes Z3 and ZJ, respectively). \\

Six stations are borehole ones, meaning that the sensor is not directly on the surface but in a borehole with various depths (five is at 150 m and one is at 75 m). Seismometers are installed at the bottom of the boreholes. Since these are not closed tight, fluctuations in the air pressure can actuate at the bottom of the borehole where the seismometer is located. There are 10 short period stations, while the others have broadband instrumentation. All three components of ground motion data are recorded at each station with a sampling rate of 100 Hz. Data centre of the national network is located in Budapest and all the stations have near real time data access. The SeisComP3 software package is used for data acquisition and processing. \\ 

\begin{table*}
 \caption{Comparisons of different meteor observational methods. The radio and spectral observations have not been discussed in this work in details, as such detectors have not been involved in the system of the authors yet. * Marks that the maximal brightness is difficult to reconstruct.}
 \label{tab:observations}
 \begin{tabular}{SMMMMLL}
  \hline
  Method & Localization accuracy & Size threshold of meteoric bodies & 3D trajectory & Compositional information & Energy estimation & Spatial and temporal coverage \\ \hline
  Optical & High (degrees on the sky) & mm & Possible to reconstruct & No & Using empirical formulae, mainly below dm size* & Only above observed (200-300 km diameter) area and nighttime \\ \hline
  Radio / radar & Moderate, mainly azimuthal directions & mm & Difficult to reconstruct & No & Not yet established, but might not be accurate & Only above observer (500-800km diameter area), anytime \\ \hline
  Ionospheric & Low (only azimuthal directions) & Not yet established & Difficult to reconstruct & Possible ablated metal atoms / ions & Not yet established & Only above observer (200-300 km diameter area), anytime \\ \hline
  Spectral & High (degrees on the sky) & cm-dm & Possible to reconstruct & Ionized components well determined & Using empirical formulae, mainly below dm size & Only above observer (200-300 km diameter area) and nighttime \\ \hline
  Infrasound & No & Above about m & No possibility to reconstruct & No & Possible, but should be much improved & Several 1000 km, for large flare ups global, anytime
 \end{tabular}
\end{table*}


The basic characteristics and comparison of various meteor observation methods are listed in Table \ref{tab:observations}. It is already known from many years of experience that some methods reached their accuracy limit (like radio based determination of atmospheric flight trajectory \citet{kozlovsky2020}). While both radio and radar observations detect ionized components of the meteor trails in the atmosphere, combined these might cover a larger part of the flight track. This would mean, that not just the ionospheric passage would be registered (which records the consequence of the deposited and produced ions, not the direct radio reflection from them). 

 \subsection{Ionosphere observation methods}
 \label{sec:ionosp}
 
\begin{figure}
    \centering
    \includegraphics[width=\columnwidth]{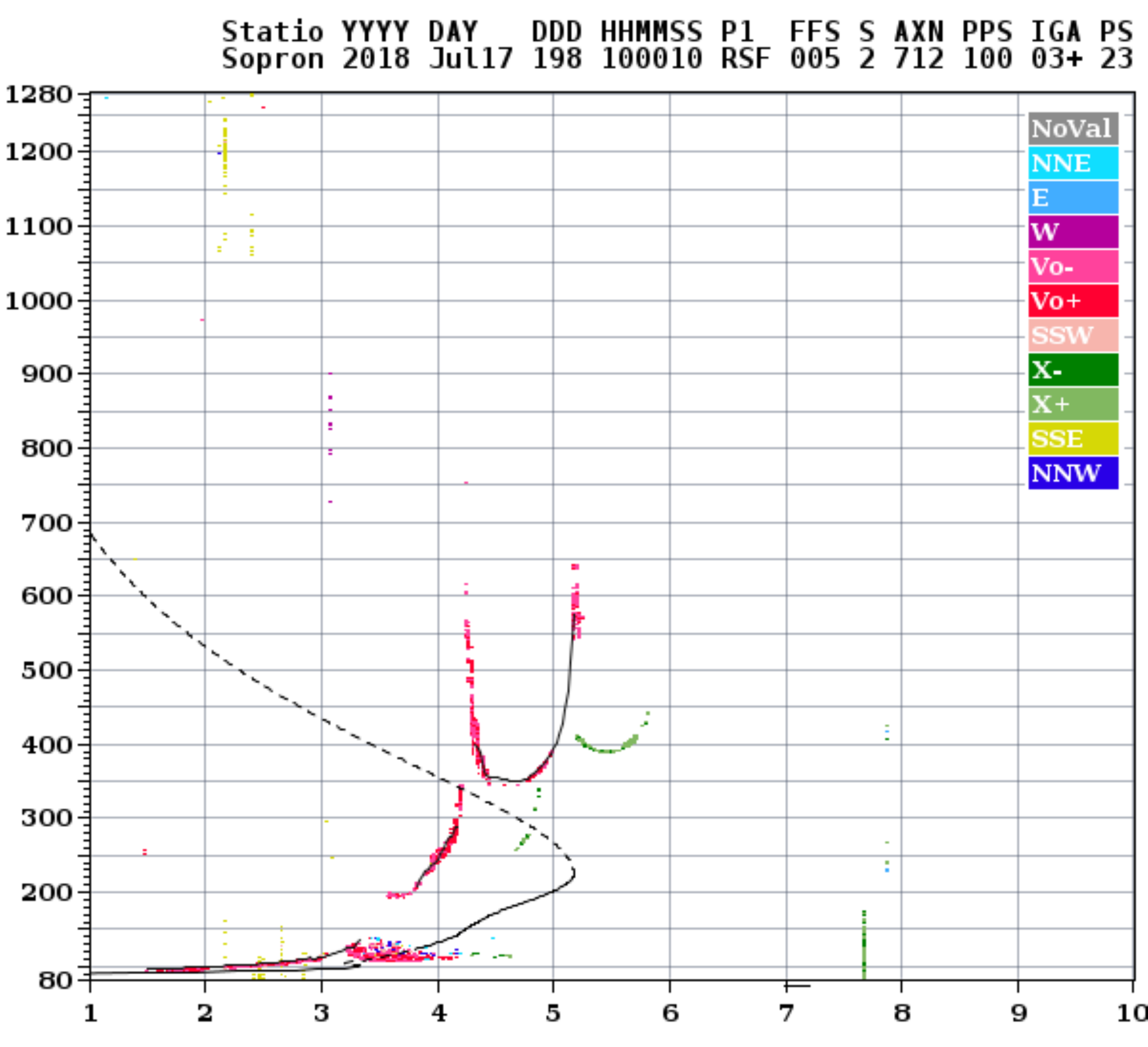}
    \caption{An ionogram detected at 10:00 (UTC) on 17 July, 2018. The different colors indicate the direction of the received signals. The different layers of the ionosphere are shown by the red/pink colored traces which are the results of the vertical sounding mode. The reflected sounding frequency (thus the plasma frequency of the layers) is seen on the X axis in MHz, while the virtual height of the reflection is seen on the y axis in km. The black curve shows the calculated electron density profile based on the measurement. The event was recorded without identifying the corresponding meteor.}
    \label{fig:ionogram}
\end{figure}

 Vertical sounding of the ionosphere is based on the reflection of electromagnetic waves from ionospheric plasma. The ionosonde emits electromagnetic pulses, which are reflected from the ionosphere when the sounding frequency is equal to its plasma frequency (e.g. \citet{davies1990}). The typical frequency range used by the ionosondes is 1-20 MHz and it provides the observation of the ionosphere from E layer up to the maximum electron density in the F region. By increasing the frequency of the sounding wave\textbf{,} the pulse penetrates into higher layers of the ionosphere. The maximum frequency of the reflected wave from a particular layer is related to the maximum plasma density of the layer -- this is called critical frequency. Above that, the pulse propagates through the layer and reflects from the neighbouring layer. Above the maximum electron density of the ionosphere the pulse propagates through it without any reflection. The result of the measurement is the so-called ionogram, representing height-frequency characteristics of the ionosphere. The layers appear as traces in the ionogram (Figure \ref{fig:ionogram}). Using the frequency-height pairs of the vertical echo, one can receive the height profile of the electron density \citep{davies1990}.\\
 
A Digisonde DPS-4D ionosonde was installed at the Széchenyi István Geophysical Observatory (47.63°, 16.72°) in the framework of GINOP-2.3.2-15-2016-00003 (titled “Kozmikus hatások és kockázatok" national project) and it has been operating since June, 2018. The ionosonde monitors the ionosphere in 15 min time resolution in standard mode. This Digisonde DPS-4D operates in a multi-beam sounding mode using six digitally synthesized off-vertical reception beams in addition to the vertical beam. The reflected beams are received by four antennas then the raw data are collected, processed and are represented for each frequency and height on a multi-beam ionogram \citep{reinisch1996,reinisch2005}. The direction of the reflected sounding pulse can also be determined due to this technique. These different colors of the traces on the multi-beam ionogram indicate the direction of the received EM pulse (Figure \ref{fig:ionogram}). 

\subsubsection{Infrasound observation methods}

The first infrasound array of Hungary was installed in May\textbf{,} 2017 \citep{czanik2017,czanik2020} and it was used in this work. The array is located in the Mátra Mountains, Northern Hungary. It consists of 4 elements, with an aperture of approximately 250 m. All the elements are equipped with a SeismoWave MB3d microbarometer with a built-in digitizer and a rosette wind-noise reduction system \citep{alcoverro2005} made of flexible hoses. The site is covered by a mixed forest, which helps to further reduce environmental noise. The central element of the array (PSZI1) is co-located with a broadband seismometer PSZ, jointly managed by the Kövesligethy Radó Seismological Observatory and the GFZ German Research Centre for Geosciences.\\

PSZI is registered at the International Registry of Seismographic Stations, and the HNIN (Hungarian National Infrasound Network) has the FDSN (International Federation of Digital Seismograph Networks) network code HN\footnote{https://geofon.gfz-potsdam.de/doi/network/HN}. PSZI is also co-located with the Konkoly Meteor Observing Network. The HNIN is a member of the Central and Eastern European Infrasound Network (CEEIN), established in 2018 by the CSFK Kövesligethy Radó Seismological Observatory, Budapest, Hungary, the Institute of Atmospheric Physics, Czech Academy of Sciences, Prague, Czech Republic, the National Institute for Earth Physics, Bucharest, Romania and the Zentralanstalt für Meteorologie und Geodynamics, Vienna, Austria. Members of the CEEIN exchange data real time and collaborate in scientific research on regional infrasound sources.\\

The CSFK Kövesligethy Radó Seismological Observatory has joined the Atmospheric dynamics Research InfraStructure in Europe (ARISE\footnote{http://arise-project.eu}) in 2016. ARISE2 is an infrastructure Design Study project funded by the H2020 European Commission. It is a collaboration of more than 25 European universities and research institutes \citep{blanc2015}. ARISE2 aims at providing a new atmosphere model with a high spatio-temporal resolution by integrating different techniques such as infrasound, lidar, radar and airglow stations. The ARISE2 infrasound network component incorporates the Comprehensive Nuclear-Test-Ban-Treaty Organization (CTBTO) International Monitoring System (IMS) network globally distributed infrasound arrays as well as several European infrasound arrays. 	 \\

PSZI detects infrasound signals from various sources including microbaroms from the North Atlantic Ocean, supersonic and subsonic aircrafts, quarry blasts, thunderstorms, volcanic eruptions and meteors. Detected events are published yearly in the Hungarian Seismo-Acoustic Bulletin \citep{bondar2019a,bondar2019b}.

\section{Results}

In this section such example results are presented, that demonstrate how some meteor observational methods could be linked (like optical-ionospheric, infrasound-seismic types). Although there was no meteor event that was observed simultaneously by all methods used in this work, the aim is to provide an overview of the potential of such level of synergy. The restricted evaluation of joint observations (usually the same meteor by two methods) provides useful examples for the rationality and ideal ways how even more complex multiple simultaneous observations can improve our knowledge -- this is discussed in Section \ref{sec:disc}. 

\subsection{Effect of meteor activity on the ionosphere}

While the correlation of ionospheric effects and specific meteors was rarely found yet, some examples below outline the possible benefit of the simultaneous observations (even for a single meteor). The ionosonde provided the opportunity to compare high cadence ionograms measured during meteor showers with simultaneously obtained optical data to determine the plasma trails of individual meteors. Campaign measurements with two ionograms/minute have been performed during the Leonid (16-18 November) and Geminid (10-15 December) meteor showers in 2019.  \\

Leonid campaign: As the sky was cloudy on the nights of 16 and 17 November 2019, the optical detection of the meteors was possible only on the night of 18 November. The camera detected only 9 meteors, possibly due to the full Moon. In the first step the high-resolution ionograms were evaluated manually to determine faint, short-lived Sporadic E layers, visible as the signal of individual meteor trails on the ionograms. The selected ionograms were compared with the optical measurements. There were only 3 cases when Es activity related to the optically observed meteors were detected. \\

\begin{figure}
     \centering
     \includegraphics[width=\columnwidth]{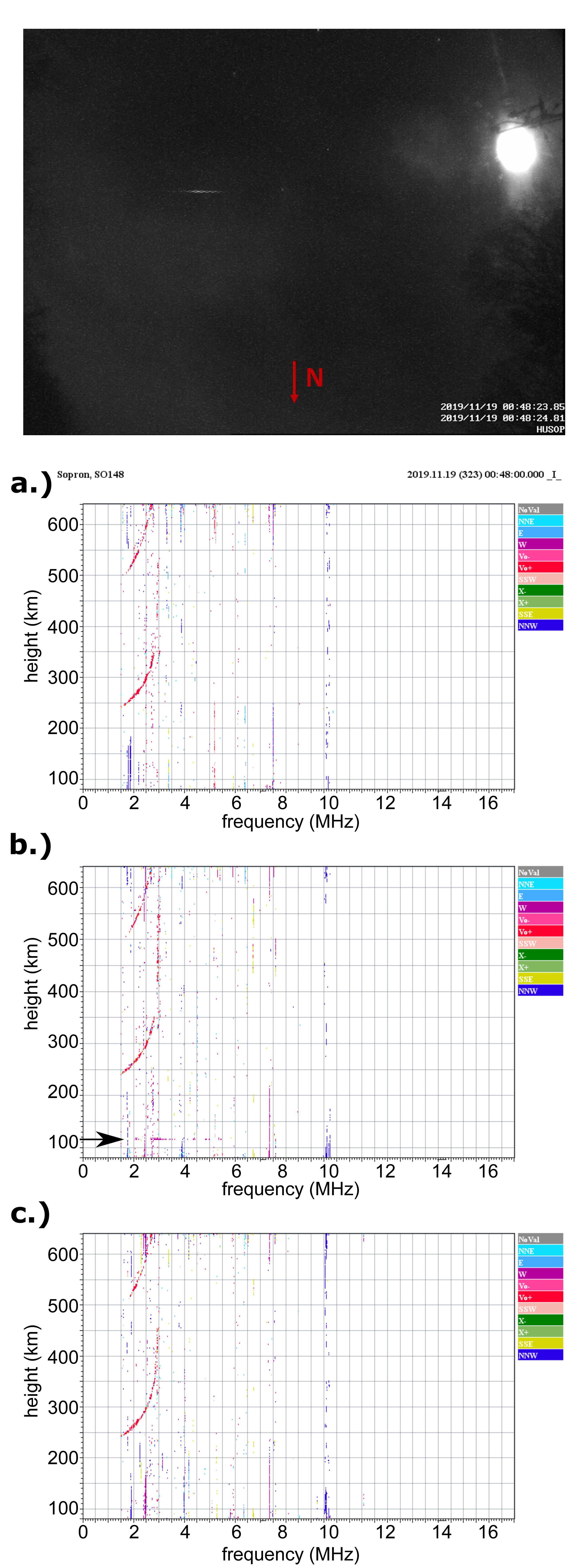}
     \caption{Top: Optical observation of the meteor at 00:48:23 UTC on 19 November 2019. Bottom: Ionograms recorded at Nagycenk station at 00:48:00 (a.), 00:48:40 (b.) and 00:49:00 (c.) respectively on 19 November 2019. The color of the detected traces indicate the direction of the received signal. The black arrow shows the detected faint Sporadic E echo.}
     \label{fig:leonida}
 \end{figure}

Here we present one of the common observations that has been recorded during the Leonid campaign. This example meteor indicated in Figure \ref{fig:leonida} was optically observed at 00:48:23 in the West-South-West direction from the zenith, with a maximum brightness between +0.9 and +1.0 magnitude at high elevation around 85° above the horizon. \\

\begin{figure}
    \centering
    \includegraphics[width=\columnwidth]{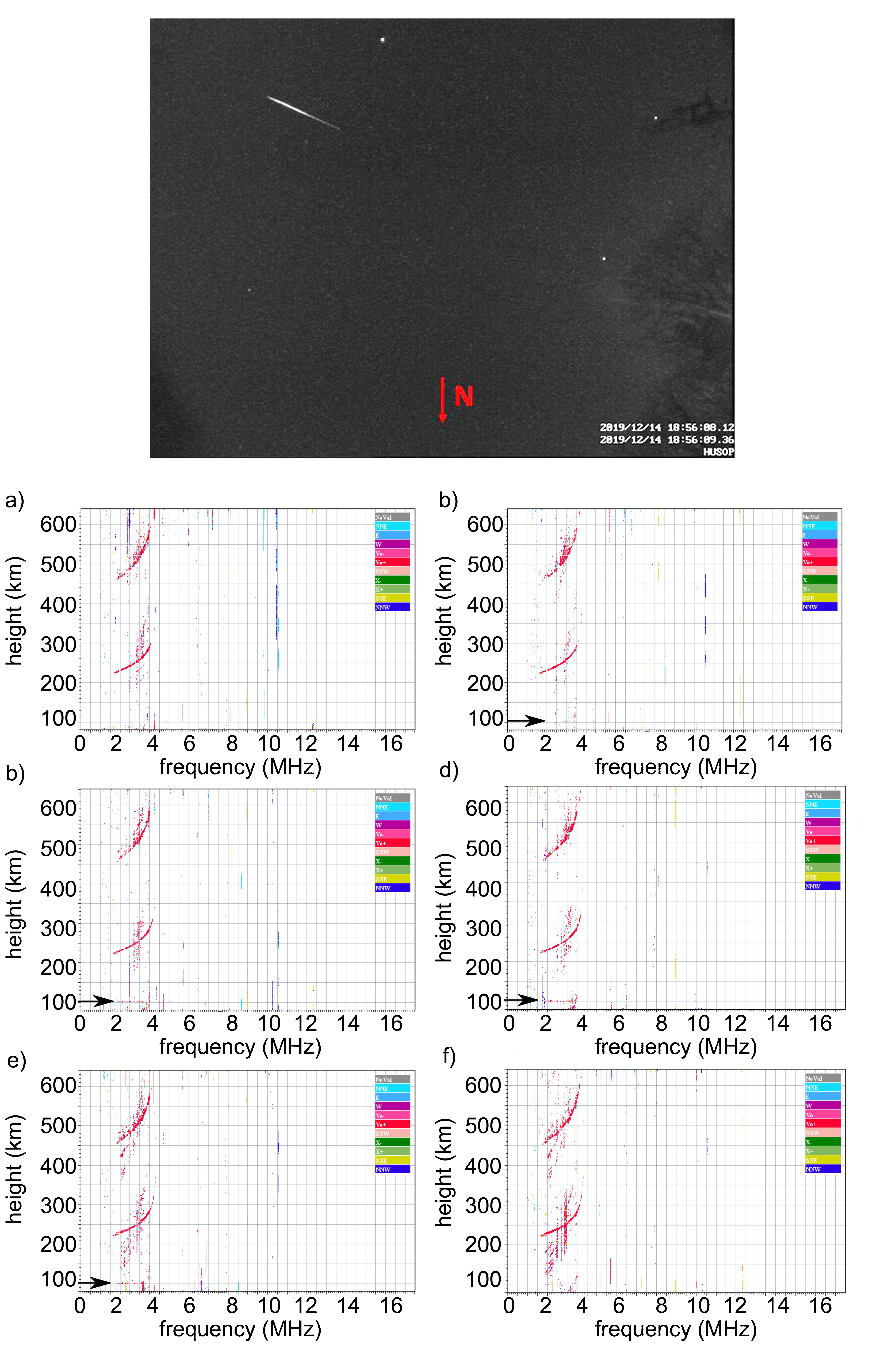}
    \caption{Top: Optical observation of the meteor at 18:56:08 UTC on 14 December 2019. Bottom: Ionograms recorded at Nagycenk station at 18:56:36 (a.), 18:57:00 (b.), 18:57:36 (c.), 18:58:00 (d.), 18:58:36 (e.) and 18:59:00 (f.) respectively on 14 December 2019. The color of the detected traces indicates the direction of the received signal. The black arrows show the observed faint Sporadic E traces.}
    \label{fig:geminid1}
\end{figure}

A faint sporadic E layer up to 5.5 MHz at 115 km height can be identified on the ionogram recorded at 00:48:40 UTC (indicated by black arrow in Fig. \ref{fig:geminid1}). There was no observed Es activity on the ionograms before and after this event (Figure \ref{fig:geminid1}). It is a good example for a faint, short-lived Es layer (max. 20 seconds), which is a typical signal of an individual meteor trail on the ionogram\textbf{, }based on previous studies. Furthermore, the direction of the echo can be also defined on the DPS-4D Digisonde ionograms thanks to the multi-beam observation technique. The direction of the detected Es layer is west (purple color on the ionogram, Figure \ref{fig:geminid1}) which agrees with the optical observation. \\

Comparing the ionograms with the closest ionosonde observation at Pruhonice station (50°N, 14.6° E) at the same time, we can conclude that the detected Es layer is a local plasma irregularity. No Es activity was observed at Pruhonice at this time of the night. This strengthens the hypothesis that the observed trail on the ionogram represents the echo of the optically recorded meteor occurred above the Nagycenk Observatory, and not an unexpected ionospheric event that emerged regionally. \\

Geminid campaign: The two ionograms/minute high time resolution campaign measurement was performed during the Geminid meteor shower (10-15 December 2019), too. The sky was cloudy on the nights of 10, 11, 12, 13 and 15 December, thus the zenith camera has captured only a couple of meteors between the clouds. However, the meteorological circumstances were favorable on 14 December. 68 meteors were recorded in total by the camera during the night. The same steps of the analysis have been repeated as in the previous case. In 18 cases short-lived (1-2 min) sporadic E layers were observed during or soon after the time of the meteor strikes. Two samples of the joint observations are detailed below. \\


The camera detected a meteor at 18:56:08 in the S-SW direction from the zenith (Fig. \ref{fig:geminid1}). Approximately a minute after the optical observation an Es layer appeared at \textasciitilde 100 km height on the ionogram (at 18:57:00 UTC Fig. \ref{fig:geminid1}, the observed layer is indicated by black arrows). The layer was observed on 4 consecutive ionograms, thus its lifetime was \textasciitilde 2 min. The direction of the received signals reflected from this short-lived plasma irregularity is S-SW (from the zenith, faint rose color on the ionogram) which agrees well with the direction of the meteor detected by the camera. \\

We compared our ionosonde observation with ionograms recorded at Pruhonice at the same time. According to the detailed comparison the Es layer observed at Nagycenk is a local plasma phenomenon, it is not detected at Pruhonice. Therefore, it is probably the plasma trail of the optically observed meteor. \\

\begin{figure}
    \centering
    \includegraphics[width=\columnwidth]{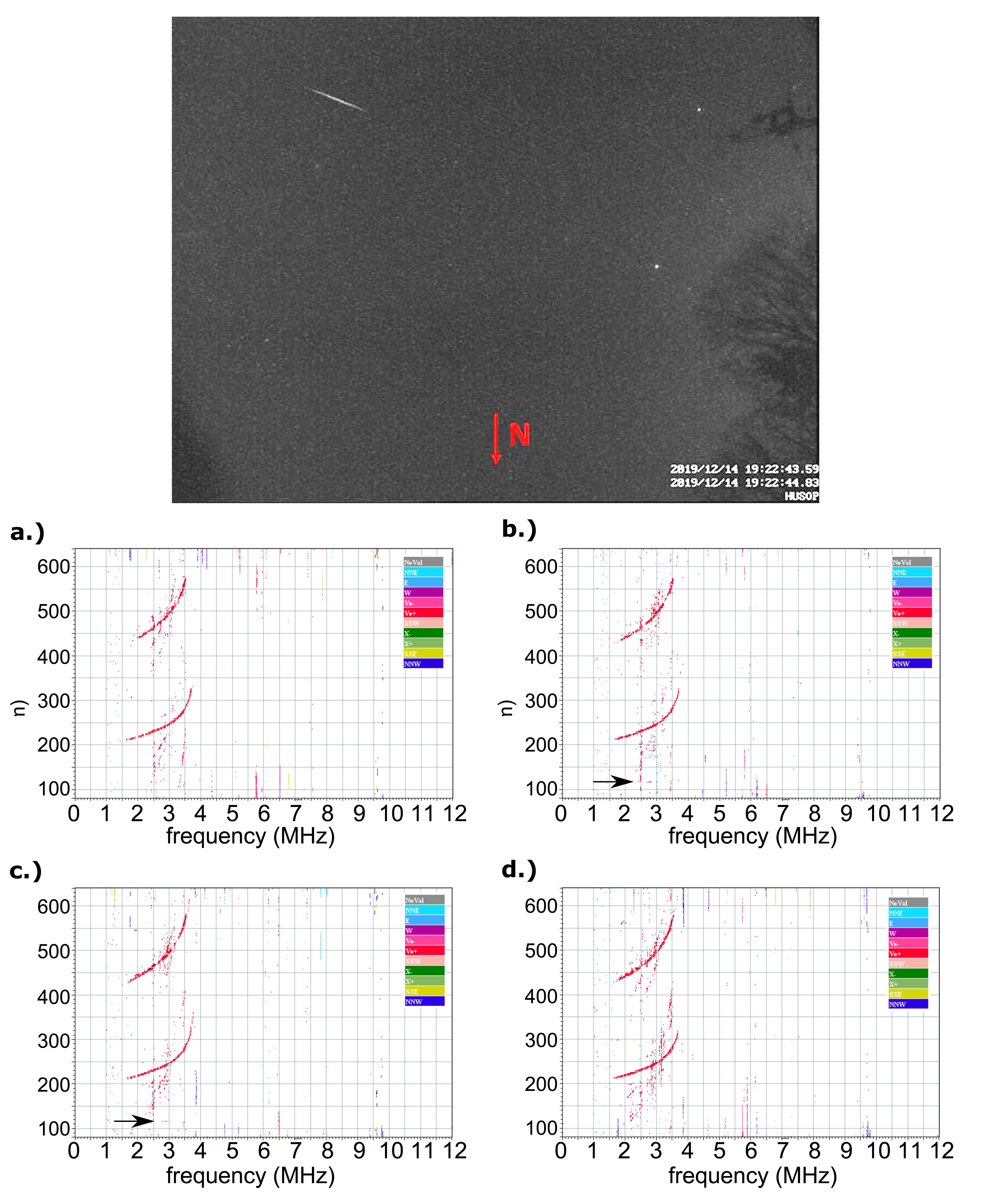}
    \caption{Top: Optical observation of the meteor at 19:22:43 on 14, December, 2019. Bottom: Ionograms recorded at Nagycenk station at 19:23:00 (a.), 19:23:36 (b.), 19:24:00 (c.) and 19:24:36 (d.) respectively on 14, December 2019. The color of the detected traces indicate the direction of the received signal. The black arrows show the observed faint Sporadic E traces.}
    \label{fig:geminid2}
\end{figure}

Another meteor was detected by the camera at 19:22:43 UTC at SSW direction from the Zenith (Fig. \ref{fig:geminid2} top). Approximately a minute later an Es layer appeared at 117 km height on the ionogram (recorded at 19:23:36, Fig. \ref{fig:geminid2}, indicated by black arrows). The Es activity was observed only on two consecutive ionograms\textbf{, thus}, the lifetime of the layer was \textasciitilde 1 min. Therefore, it is a faint, short-lived Es layer which can be considered as\textbf{ a} plasma trail of a meteor according to the literature. The direction of the echo is S-SW from the zenith based on the color code (faint rose), therefore it agrees well again with the direction of the optically observed meteor. Instead of the short-lived thin layer\textbf{,} a regular strong Es layer is detected at Pruhonice at 19:23-19:25 UTC time interval and before/after.  This seems to reassure that the observed Es activity detected at Nagycenk is related to a local plasma irregularity, probably caused by the meteor. 

\subsubsection{Infrasound and seismic effect of meteor activity}

Around the Croatian-Slovenian border region a large bolide could be seen on the daytime sky on 28 February 2020 \citep{ott2020}. A 1.5 m diameter meteoroid entered the atmosphere at around 9:30 UTC, and probably fragmented catastrophically at the height of cca 34.5 km above the ground roughly 75 km west of the city of Zagreb releasing 0.34 kt equivalent energy by its flare up \citep{carbognani2020}. A lot of people from SE Slovenia alarmed the Slovenian Seismological Agency reported that have felt an earthquake. Also, the European-Mediterranean Seismological Centre received a lot of earthquake reports from Zagreb, Croatia. But after reviewing the seismic records it was obvious that it did not show the characteristics of seismological events (details can be found in Figure \ref{fig:idosor} below). \\


\begin{figure}
    \centering
    \includegraphics[width=\columnwidth]{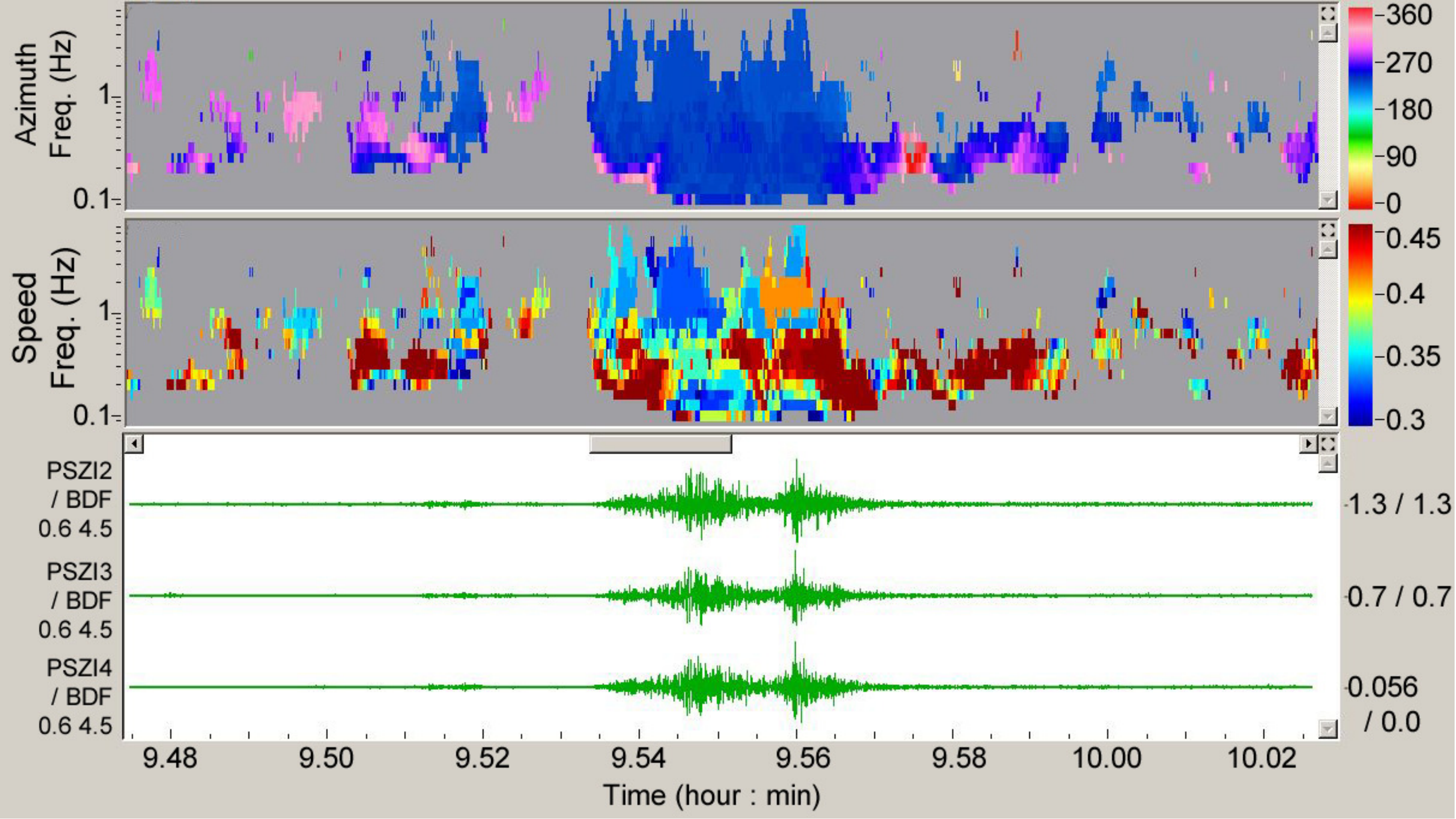}
    \caption{The image of the registered infrasound (top) and the effect according to three main directions (bottom)}
    \label{fig:infra}
\end{figure}

Figure \ref{fig:infra} shows the Progressive Multi-Channel Correlation (PMCC; \citet{cansi1995}) results of the detection of the bolide. The lower panel shows the infrasound waveforms at the array elements filtered between 0.6 and 4.5 Hz. The middle and top panels show the time-frequency plot of the record, color-coded by the apparent speed and azimuth, respectively. The signal comes from the same direction, from about 230$^\circ$ azimuth, but it may also exhibit multipathing, as the first reflection from the atmospheric boundaries travels with about 0.33 km/s, the second reflection sweeps through the array with a somewhat larger, 0.4 km/s apparent velocity. \\

Based on the reports and the seismic records, Slovenian seismologists estimated the path of the meteoroid to have a NW heading, and the epicentre of the blast had coordinates of 45.9N 14.9E \footnote{Mladen Zvcic personal communication}. \\

In spite of the fact that it was an atmospheric event, its traces could be seen on seismic records, too. (Remember, seismometers installed in seismological stations measure ground movement.) It is quite common that seismometers can record the acoustic traces of nearby catastrophic fragmentations. It is because the shock wave of a nearby disruption can agitate the ground and this ground movement can be recorded by the seismometer. But generally this acoustic part of the signal attenuates quite fast.

\begin{figure}
    \centering
    \includegraphics[width=\columnwidth]{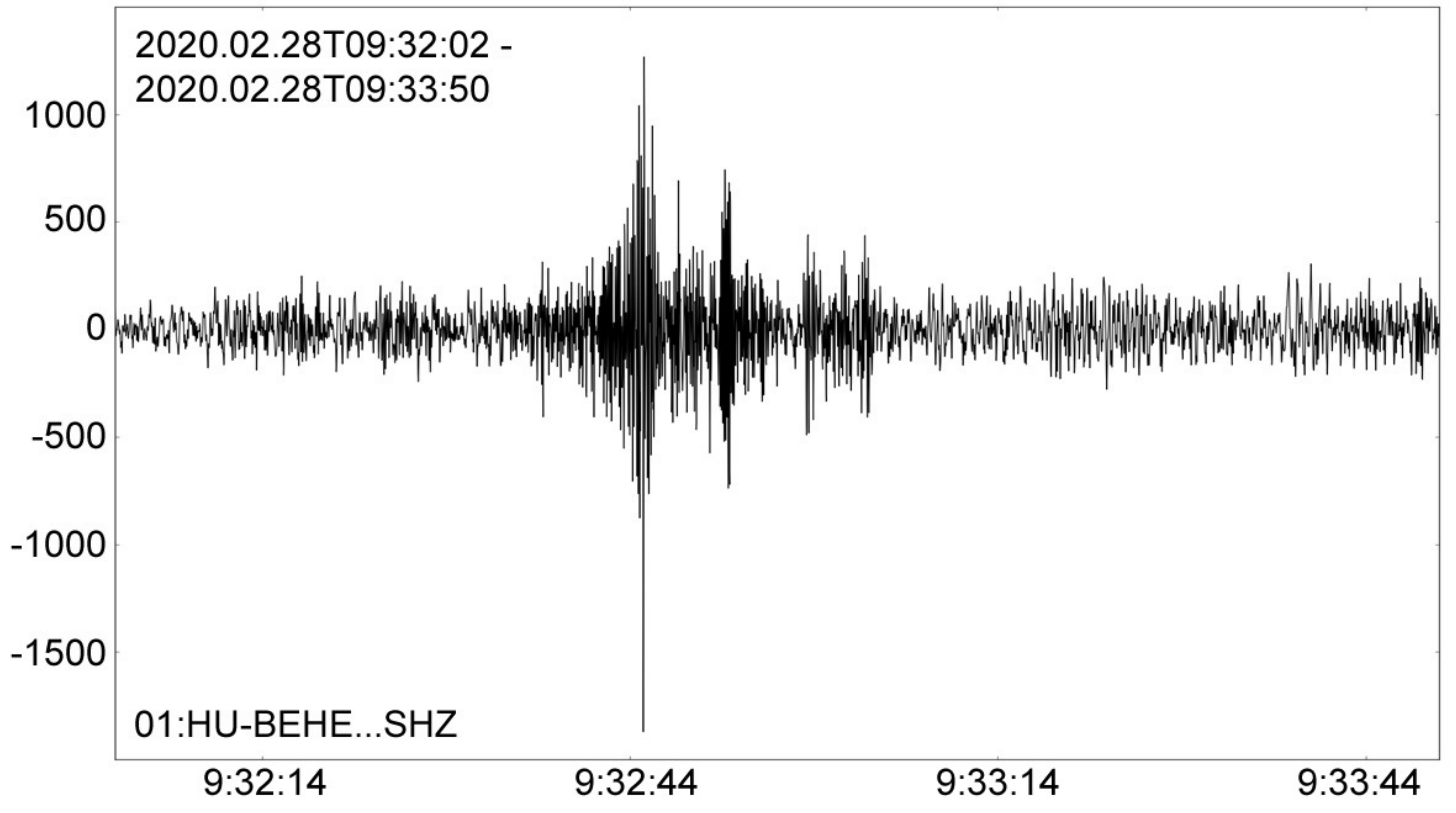}
    \caption{Vertical ground movement at station BEHE (Becsehely, Hungary) after bolide \textbf{flare up} at a distance of 160 km (duration of this section is 1.5 min)}
    \label{fig:behe}
\end{figure}

The nearest Hungarian seismological station at Becsehely (station code: BEHE) showed a marked signal on the seismogram but it was clearly of no earthquake origin (Figure \ref{fig:behe}). With an earthquake, in the classical case P waves arrive first, S waves second and surface waves even later and these could be identified as separate features. Then the records of all Hungarian stations (except project stations) have been examined. Arranging the seismograms in the order of increasing distance from the epicentre of the fragmentation one could discover a disturbance in the record of almost every station, even in the case of borehole ones (Figure \ref{fig:idosor}). The nearest Hungarian station (BEHE) had an epicentral distance of 160 km, the farthest was TRPA (Tarpa) with 630 km. \\

\begin{figure}
    \centering
    \includegraphics[width=\columnwidth]{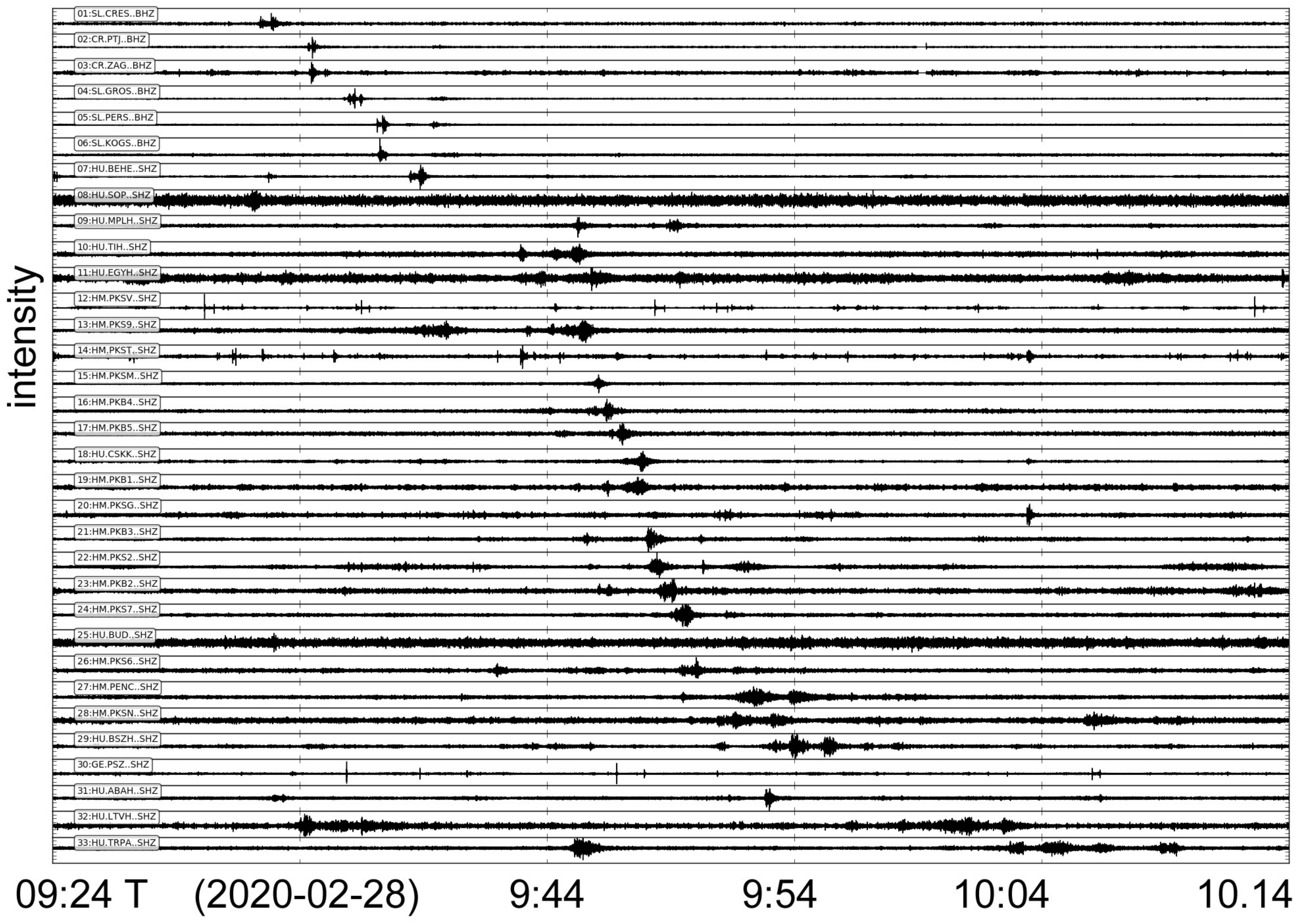}
    \caption{Fifty minute sections of seismograms from seismological stations in the order of epicentral distance. The upper 6 traces are from Slovenian and Croatian stations, all others are from stations of Hungarian networks.}
    \label{fig:idosor}
\end{figure}

Disturbance sweeping across the station network can be seen \textbf{at} almost all stations. By measuring the arrival times of this signal at each station, the apparent speed of the disturbance can be calculated (the speed as the disturbance swept across the network) (Figure \ref{fig:idosor}). According to our calculation the apparent speed of this signal is 312 m/s\textbf{,} which is very close to the atmospheric sound speed. This suggests that the shock wave in the air generated by the catastrophically fragmenting bolide was strong enough to be recorded by seismometers even at a distance of several hundred kilometers. \\


\begin{figure}
    \centering
    \includegraphics[width=\columnwidth]{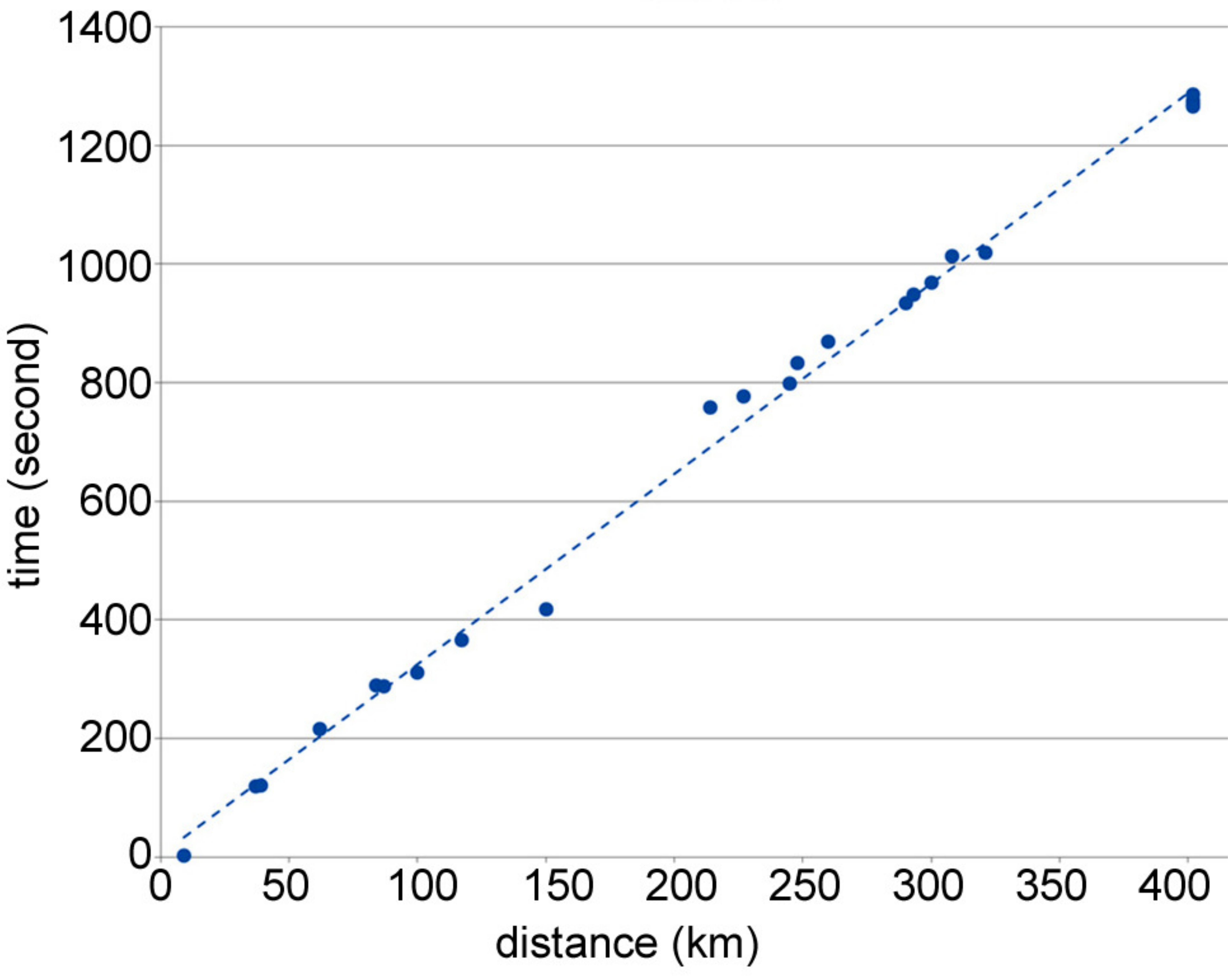}
    \caption{Travel times of the disturbance measured at different seismic stations vs. epicentral distance. Slope of the fitted line shows the apparent velocity of the shock wave which is 312 m/s in this case.}
    \label{fig:traveltime}
\end{figure}

There were some stations where this signal could not be observed. At the SOP (Sopron, 210 km) and BUD (Budapest, 320 km) stations, the location of the sensors can explain this non-detection, as both have a rather long vault and the seismometer is installed deep under the ground surface. Because of multiple reflections from atmospheric layer boundaries and mountain ranges, as well as multipathing, it becomes difficult to resolve the ablation history of this event. The seismic record does not show clear evidence of subevents (i.e. several \textbf{flare ups }during the break-up of the bolide). 

\section{Discussion}
\label{sec:disc}

In the last decade, several meteors have been observed by multiple methods (optical, radar, ionospheric etc.), however, not one of them have been observed by all types simultaneously. As a result, certain aspects could be identified \textbf{of} how multiple observation types could support the better understanding of the meteor event and the local environmental (atmospheric) conditions. These methods record different consequences of meteor activity with complementarity of identifying meteors and their characteristics. Most meteors are recorded by optical cameras in nighttime, however the optical method does not observe daytime phenomena excluding the very few extremely bright bolides and those that occur during cloudy nights. Below the comparison of the benefits and weaknesses are presented for the surveyed observational types. \\

\subsection{Ionospheric effects}

Generally, observation of the meteor plasma trails on ionograms is difficult since the ionosondes operate in 10-15 min time resolution in the standard mode. Thus rapid run ionosonde measurements were performed during meteor showers \citep{goldsbrough1976, maruyama2003, maruyama2008} however the ionograms have not been compared with local optical meteor records before. There is an obvious knowledge gap in the identification of ionospheric effects linked to specific meteors, and increased ionospheric sampling frequency. Therefore, the installation of the zenith optical meteor camera next to the DPS4D Digisonde at the Széchenyi István Geophysical Observatory in our project provided an exceptional opportunity to observe and determine plasma trail of meteors with one to one comparison of the optical observations on high cadence (\textasciitilde 30 sec) ionograms. \\

Short-lived faint Es layers have been determined during or after the optical observation of meteors, in the case of roughly the third of the optically observed meteors. These observed traces during the Leonids and Geminid showers can be defined as type 1 Sporadic E layer (Em) based on the \citet{goldsbrough1976} classification. The correlation between the height and the frequency of the individual observed layers is weak, thus each recorded event is spontaneous and it seems to represent ``backscattering from the region of increased electron density along the meteor path, which acts like a metallic cylinder'' \citep{maruyama2003}. On the contrary, there is a strong correlation between the parameters (frequency, height) determined from consecutive ionograms, thus the echos are reasonably stable in the case of normal stratified Sporadic E layers \citep{haldoupis2011}. Furthermore, the observed echos of meteor trails are weaker compared with the typical ones from the sporadic E and F layers, which also agrees with the results of the previous studies \citep{maruyama2003,maruyama2008}. \\

By using four receiver antennas and the multiple beams technique of the DPS4D Digisonde \citep{reinisch1996,reinisch2005}, we were able to define the directions of the sounding pulse reflected from the spontaneous Es layers. The determined directions agreed well with the directions compared to the zenith of the meteor in the optical observation in all cases. We compared the observed ionograms with the ionosonde measurements at the closest station (Pruhonice, located at \textasciitilde 200 km far in N-NW direction from Nagycenk) during the investigated periods. Most of the optically observed meteors and the recorded echos were in S-SW direction from the zenith. The visual angle of a multi-beam Digisonde is $\pm$45$^\circ$ from the zenith (Digisonde manual link), which means a \textasciitilde 100 km radius at the height of the Es layers (100-120 km altitude). Therefore, the observed meteors were too far from Pruhonice station to detect them by the Digisonde installed there. \\

We detected the meteor induced spontaneous Es layers in several consecutive ionograms, thus the lifetime of these layers ranged between 1-3 min. This agrees with the duration of the meteor trails observed by high cadence ionosonde measurements during the Leonids 2001 meteor shower \citep{maruyama2003}. However we did not observe any meteor induced long-lived (>20 min) Es patches reported by \citet{maruyama2003} and \citet{maruyama2008}. It is possible that much more meteor phenomena have left Es layers but because of the moderately long sampling interval they have gone undetected. 

\subsection{Infrasound observations}

Infrasound generated by the daytime fireball on 28 February 2020 was recorded, determination of azimuthal direction was possible. Signatures of multipathing suggest complex interaction with various atmospheric layers, and ground shaking events were observed by several related seismological stations in Hungary. \\

Infrasound based detection provides the following complementary aspects relatively to other observing methods. While optical and ionospheric observations could be made only at 100-200 km distance from the location of the atmospheric event, seismic effect could travel to 1000 km scale (in the case of >10 m size object), and infrasound produced by roughly >1 m objects could travel even farther away (several 1000 km distance). Thus 1-10 m sized objects might be observed by infrasound from the largest distances, even larger than by radio methods (excluding satellite based data). In an ideal case, knowing the distance of the event, infrasound detection could provide mass estimates calculated from the blast-wave theory of \citet{revelle2008b} -- thus being complementary to optical detection (for example oversaturated images), and also improve the understanding the strength of the body (as this later influences the height of the flare up together with the size/mass of the object). The opposite is also relevant: having space based optical or infrared data on atmospheric entry and breakup \citep{mcintosh1976}, the method to calculate the released energy could be improved by infrasound based estimation. A further benefit of infrasound detection based fireball observing is that it is independent of the state of the sky (could be day- or night-time, cloud covered or even stormy period). However it should be also considered that multiple reflections of various atmospheric layers and topographic features with specific geometry might provide multiple reflections for infrasound which should be filtered by improved methods in the future. \\

Infrasound data facilitate the localization of the meteor event. The azimuth measurements of infrasound stations are typically so accurate that the direction of the meteor can be determined based on this observation alone. In the case of observations from several different locations, the spatial determination might be even better. However, relying only on the azimuth observations will not provide the height of the catastrophic fragmentation; for that arrival time observations are also needed. In most cases a simple celerity estimation is used, but that significantly increases the location uncertainties as the accurate velocity distribution in the atmosphere can only be described by a 4D model. More sophisticated studies use ray-tracing through the atmosphere by models from European Centre for Medium Weather Forecast Period (ECWMF) or \textbf{the} National Oceanic and Atmospheric Administration (NOAA). The estimates for the total energy release based on infrasound observations are derived from atmospheric nuclear explosions of 10 kt or larger events (roughly produced by the atmospheric fragmentation of a 10 m sized meteoric object). Below that the uncertainty in the estimated energy release increases, but for smaller objects optical based estimation is moderately accurate (as the optical detectors do not get saturated by smaller bolides). 

\subsection{Seismological observations}

Although the appearance of the seismological signal of the atmospheric blast differed from the characteristic seismological events, it is poorly known what characteristics of the bolide can be decoded from it. Using the observation of the above event it turned out that the transfer of an airblast to a seismic event depends strongly on local lithological conditions, and its parameters are difficult to determine using seismological signals. However, it demonstrates that 1.5 m category objects breaking up 30-40 km above the ground can be easily detected with many stations. \\

Based on earlier results several atmospheric fragmentation events could be identified occasionally \citep{atanackov2009}. An important work to be done here is the clarification of the connection between atmospheric effects and the triggered seismological consequences. The latter depends highly on the local lithological and surface topographical conditions, as the atmospheric blast interacts specifically with the surface being converted to seismological movement of the solid rock units by a local site dependent transfer of the atmospheric effect \citep{karakostas2015}. For example the Chelyabinsk event produced an earthquake of 3.85 magnitude, while the Tunguska event of 1908 resulted in a 4.8 – 5.0 magnitude earthquake \citep{svetsov2014}.  \\

Important joint observations were made in 2007 in the region of the Bolivian / Peru border on 15 September, related to a bright fireball \citep{brown2008,borovicka2008,tancredi2009}. An impact crater finally called Caracas was also discovered to the south of Lake Titicaca, produced by this event. Beside the detection by seismic networks, two infrasound stations have also observed\textbf{ it} at roughly 80 and 1 620 km from the crater. During the atmospheric flight, shock waves were generated along a cylindrical line emission and also from point-like flare up. Data interpretation showed that the dominant process in the sound generation was the fragmentation, rather than the hypersonic shock during the flight. The two strongest sound effects were generated at 25 and 35 km altitudes with energy release equal to about 1.0 and 3.8 t TNT, respectively. From the crater diameter the impact released 4.8 t TNT. The impact location found by the seismic observations coincides with the crater location. From the crater diameter and the airwave amplitudes, the kinetic energy, mass and explosive energy were accurately calculated. \\

Future attempts on the systematic search for seismological consequences of airblasts from meteoroids would be extremely important. The signatures might be moderate or small, but with the joint evaluation of seismograms from several stations, it could be possible to identify several events that would have gone unnoticed earlier, especially as seismological networks are more densely arranged than infrasound networks on Earth. Seismological data archives should be surveyed in the future at the dates of large fireballs. 

\subsection{Suggestions for further improvements}

There is a wide range of possibilities for the identification of connections and discoveries with the joint evaluation of different meteor observation types. Different threshold limits of observations exist regarding bolide size, brightness, spatial and temporal occurrences of meteors for the synergy of the presented methods. The smallest meteoritic bodies (down to cm size) could be identified optically, while the effect of meteoritic bodies of this size could be also recorded in theory by ionospheric observations -- especially if there are many very faint meteors below the limit of optical detection during a short outburst period of high meteor rates. The infrasound effects and related seismic effects should be identified from larger objects (about 1 m diameter), probably in more cases than have been achieved up to now. \\

The methods discussed above are subject to continous development, and substantial technological improvements are expected of the next years. Thus it is worth to evaluate which aspects and directions should be in the focus of developments (Table \ref{tab:summary}). \\

\begin{table*}
 \caption{Summary of the observational potential of different methods.}
 \label{tab:summary}
 \begin{tabular}{MLLXX}
  \hline
  Influenced region & Specific phenomena & Observable characteristics & Subtopics to be improved & Future results can be gained \\ \hline
  Upper atmosphere & Ionization, shock wave propagation, chemical change, material input from porous meteoroids (like IDPs) & Echo of the plasma trail of individual meteors and meteor induced Es patches & Gain higher temporal resolution of ionospheric observations, link ionospheric to radio based observations, determine the lower threshold limit of events & Identification of ionization properties, estimation of ionic inputs to the atmosphere \\
  \hline
  Lower atmosphere & Atmospheric shock wave and chemical changes, free fall (dark flight) & Travelling, reflected and refracted shock waves, falling meteorite (observable by radar) & Improved noise filtering, improved interpretation of complex acoustic effect & Atmospheric flare up of meter size objects, improved statistics of small NEOs \\ \hline
  Surface & Shock wave & Seismic shock from atmospheric blast (and direct impacts), fallen meteorite & Improved search algorithms for the identification of atmospheric blast induced seismic shocks & Linking between atmospheric blasts and shock waves in the solid surface, extrapolation for larger, historical events\\
 
 \end{tabular}
\end{table*}

Evaluating Table \ref{tab:summary}, improvement of NEO statistics is expected in the range of 1-10 meter sized objects, which are almost impossible to identify by telescopes before the atmospheric light up. It is also expected that the joint evaluation of various observations provides a better coverage of the meteor events. catastrophic fragmentation of meteoric bodies above the oceans including daytime hours, where and when optical monitoring is impossible (except for a few cameras installed at certain islands, but these are only for nighttime recording, and this observation type can only cover events not farther away than 100-200 km horizontal distance). The ionospheric effects could be observed in daytime too, possibly covering twice as much meteors as with nighttime observations only -- however the difference between the behavior of nighttime and daytime reactions of the ionosphere is not well known yet.  \\

New information could be found on the ionospheric material input, with the linking of optically observed meteor speed and size to ionospheric consequences. The ionospheric ablation can be connected to the structural properties of the meteoroids, as well as the infrasound based characteristics of any terminal fragmentations (this latter is difficult to analyze using optical detectors daytime, while during the nighttime it could get oversaturated, increasing the uncertainty). \\

In spite of the previous observations, there are still open questions in the relationship between the meteor rates and Es activity, for example: the effect of meteorite showers to the sporadic E activity; the observable length of individual meteors (or meteor-induces Es patches) on ionograms; the role of the meteor shower's origin and composition (metallic ion concentration) in its effect on the Es layers. \\

Correlation between the occurrence of Sporadic E layer and optical observable meteors is expected to be quantified in the future. Recent results suggest that there is a connection not only with the occurrence of sporadic E layer in the ionosphere and meteor activity, but the composition of this layer and expected meteoroid input from space. The ablating meteors deposit a range of metallic ions there including Fe$^+$, Mg$^+$, Si$^+$, Na$^+$, Ca$^+$, K$^+$, Al$^+$. The average values of foEs are proportional to the velocity V of meteors \citep{alimov2016}, and the critical frequency of the sporadic Es layer, which increases with V, demonstrating the influence of meteor particles on ionization and formation of long-lived metal atoms M and ions M$^+$ of meteoric origin. Increased metal ion density increases ionization and electron density of the given zone in the ionosphere -- however the wind shear can complicate the situation. Even so, some compositional or structural characteristics can be roughly estimated in the future, especially if spectral meteor observations might be also integrated to the system. Differences between the optical spectra of various meteoroid streams have been also emerged from spectral observations of meteors recently \citep{ferus2020,koukal2016}, but more detailed information is required to get an insight into the composition of meteors using this approach. In the future, increased metal ion occurrence is expected to be identified, associated to meteors having increased metal content \citep{kasuga2005} -- however differences only in the Fe/Mg/Na ratio have been deeply analysed \citep{matlovic2020} yet. The light up height of meteors might have some dependence on the speed and possibly also on its structure and composition \citep{lukianova2018}. As a result the most fragile meteoritic particles (like interplanetary dust particles, IDPs, \citet{genge2020}) might provide ionospheric consequences without optically observable consequences -- which should be tested in the future. Ionospheric observations also indicated that solar activity increases the meteor ablation height, and during the solar maxima meteor peak detection height rises \citep{premkumar2018} -- the background of this process will be better understood with more ionospheric meteor observations. It is an interesting question whether this effect can influence the light up height of optically observable meteor events. If the occurrence of sporadic E layer could be better connected to the optically identified meteors, and a general understanding allows to separate the meteor produced sporadic E component on the ionograms from the regular sporadic E background, a new possibility emerges for daytime meteor observations. \\

There are several topics in the atmospheric entry and consequences of meteor events where joint analysis of different observation types is necessary. While the consequences (including surface damage from atmospheric blast) of events by 1-10 m size objects can be well predicted, the consequences of 100 m scale ones are less known as they depend strongly on the internal properties of these objects \citep{aftosmis2019} (here multiple breakup is an important aspect).  \\

The energy deposition of breakup events during the flight strongly depends on the fragmentation, internal strength and fragmentation height, which could be simulated properly if more information is gained from observations comparing ionospheric, optical and acoustic consequences \citep{robertson2015}. The optical light curve together with the infrasound data helps to determine how sudden / prolonged and single / multiple was the fragmentation, which is difficult to decode from the oversaturated optical images alone. Even some aspects of the meteoroid’s internal properties could be estimated using the characteristics, height and even multiplicity of flare ups \citep{revelle2003}, as the more fragile (cometary like composition) bodies are expected to fragment catastrophically at high altitudes even if they are the same size as rocky objects \citep{revelle1997}. Connecting the infrasound airblast duration with the optical based flight trajectory, it could be estimated if it was an extended duration burst from a suddenly fragmenting (thus stronger), or a gradually fragmenting (thus weaker) internal structured body; and as specifically, how a low angle trajectory in the atmosphere( like in the case of the Chelyabinsk event \citep{boslough2013}) influences the flare up \citep{collins2009,janches2020}.  \\

 Based on recent events, especially the Chelyabinsk one, 10-20 m size bodies could reach the Earth undetected under certain orbital conditions -- and this might be true even for 50-100 m sized bodies. However, the atmospheric terminal catastrophic disruption of such bodies could be detected by infrasound systems around the globe even during stormy conditions at the atmospheric entry site. In the case of three suitable stations, the location of the fragmentation could be determined and the energy roughly estimated. The infrasound detection can be especially useful in cloudy weather, inhibiting all optical and infrared detection, as cloud coverage varies between 0.03 and 0.3 over the globe \citep{stubenrauch2013} in general (thus resulting in an uncertainty or blink window in the observation of large bolides). Successful daytime observations have also demonstrated usefulness of the method in identifying meteors, like the CTBTO network \citep{pichon2013}.\\

Seismic observations are strongly linked to infrasound observations, and will help in the identification of blast events during thick cloud coverage. The interaction between the atmospheric blast wave and the elastic seismic waves generated in the ground is poorly known, but if joint observations on the height and value of terminal energy releasing disruption events together with their seismic reactions were both available, their evaluation would clarify several aspects of energy transfer from atmospheric blast to seismic waves \citep{svetsov2014}. Based on the above, a better correlation between the atmospheric and seismic effects would improve the understanding of the airblast energy, as it can be more accurately estimated from the seismic effect than solely from the infrasound effect. 

\section{Conclusions}

This work gives an overview on some synergic possibilities of several different meteor observational methods with presenting case studies and considering near future perspectives in exploring their joint analysis. The used meteor data were acquired by optical, ionospheric, infrasound and seismic detectors. Nearly third of the observed cases of ionospheric reactions to specific meteor events above about 0 magnitude were identified, confirming the very few earlier results that such consequences could be found between individual meteors and sporadic-E layers. No long lasting (>20 min) Es patches were identified. Ionospheric observations should be explored in the short cadence temporal domain, and metal content and other characteristics of matter properties can be estimated in the future. Using the experiences of the observed cases, installation of two optical meteor cameras at a distance of about 20-40 km from each other, both targeting the atmospheric region about 100 km above the ionosonde will provide brand new results on the upper atmosphere by determining the exact height characteristics of the observed meteor. There is an obvious knowledge gap in the identification of ionospheric effects linked to specific meteors, which can be explored in the next decade using the expected technological improvement. \\

Infrasound effects of some meteors from several hundred km distance could be also identified. In a moderately close meteor event (which was optically also identified from Hungary, where the detector system is located), the atmospheric blast and the related seismic events have also been observed by 26 ground stations. The recorded signal differed from typical seismological signals and were influenced by the local neighborhood of the detector, and in a few cases, the signal did not emerge at all as the seismometer was too deep below the surface. \\

There is an opportunity to exploit the complementary aspects of the listed meteor observations, especially in the better coverage of the occurred meteor events, as ionospheric and infrasound observations could be realized through clouds and even during daytime. Infrasound observations provide a unique possibility for the better identification of 1-10 m category objects, even from several thousand km distance -– while the daytime optical observations are biased by clouds and the terminal catastrophic disruption of bolides could oversaturate detectors. Improved infrasound observations can support the estimation of the total energy release. 

\section*{Acknowledgements}

This work was supported by the following grants. The meteorite impact rate and NEOs related aspects were supported by the NEOMETLAB project, related small bodies aspects were supported by the UNKP project. The installation of the DPS-4D ionosonde, the optical meteor observation and the synergic considerations were supported by the GINOP-2.3.2-15-2016-00003 “Kozmikus Kockázatok” project. The seismic analysis was supported by the NKFIH K128152 (“Term\'eszetes \'es mesters\'eges eredetű esem\'enyek elk\"ul\"on\'it\'ese szeizmikus \'es infrahang adatok egy\"uttes anal\'izis\'evel”) project. The infrasound analysis related aspects were supported by the bilateral agreement between the Czech and Hungarian Academy of Sciences, NKM2018-10 (``Infrasound studies in Central Europe``, PIs: István Bondár, Tereza Sindelerova). The ionosphere related aspects were supported by the bilateral agreement between the Czech and Hungarian Academy of Sciences, NKM55/2019 (``Multiinstrumental investigation of the midlatitude ionospheric variability``, PIs: Veronika Barta, Petra Koucká Knížová).

\section*{Data Availability}

Data will be made available upon request.



\bibliographystyle{mnras}
\bibliography{Kereszturietal_Meteor_R2} 








\bsp	
\label{lastpage}
\end{document}